\documentclass[pdftex,numberedappendix,appendixfloats,apj]{emulateapj}

\usepackage{multirow}
\usepackage{amsmath}

\begin{document}

\shorttitle{Dwarf galaxies under ram-pressure}

\shortauthors{Williamson \& Martel}

\title{Chemodynamics of dwarf galaxies under ram-pressure}

\author{David Williamson}
\affil{Department of Physics \& Astronomy, University of Southampton, Southampton, SO17 1BJ, United Kingdom}
\email{d.j.williamson@soton.ac.uk}

\author{Hugo Martel}
\affil{D\'epartement de physique, de g\'enie physique et d'optique, Universit\'e Laval, Qu\'ebec, QC, G1V 0A6, Canada}
\affil{Centre de Recherche en Astrophysique du Qu\'ebec, QC, Canada}

\begin{abstract}
By implementing a dynamic wind-tunnel model in a smoothed-particle chemodynamic/hydrodynamic simulation suite, we have investigated the effects of ram pressure and tidal forces on dwarf galaxies similar to the Magellanic Clouds, within host galaxies with gas and dark matter halos that are varied, to compare the relative effects of tides and ram pressure. We concentrate on how the distributions of metals are affected by interactions. We find that while ram pressure and tidal forces have some effect on dwarf galaxy outflows, these effects do not produce large differences in the metal distributions of the dwarf disks other than truncation in the outer regions in some cases, and that confinement from the host galaxy gas halo appears to be more significant than ram pressure stripping. We find that stochastic variations in the star formation rate can explain the remaining variations in disk metal properties. This raises questions on the cause of low metallicities in dwarf galaxies.
\end{abstract}


\section{Introduction}

According to the observed mass-metallicity relation \citep{2004ApJ...613..898T}, the lowest-mass galaxies have the lowest metallicities. These dwarf galaxies should be strongly affected by interactions with more massive galaxies, which could affect the slope and scatter of the mass-metallicity relation at the low-mass end. Although the role of interactions on the morphology of dwarf galaxies has been thoroughly explored \citep[e.g.][]{2003MNRAS.345.1329M,2004MNRAS.352..363M,2005MNRAS.364..607M,2010MNRAS.405.1723S,2011ApJ...740L..24K,2014ApJ...780..119K}, the {\em chemodynamical} effects of interactions on a dwarf galaxy have not been as directly or thoroughly examined \citep[although see][]{2013MNRAS.436.1191T,2018MNRAS.474.2194E} .

We have examined the role of tidal effects on the chemodynamical evolution of dwarf galaxies in a previous paper \citep[][hereafter Paper II]{2016ApJ...831....1W}. We expected that tidal stripping would preferentially remove high-metallicity outflows and thus act as a drain of metallicity from dwarf galaxies. However, we found that tidal stripping can actually {\em enhance} the metallicity of our simulated galaxies. A metallicity gradient is produced by ongoing centrally-concentrated star formation, so that when the outer low-metallicity regions of the dwarfs are stripped, the high-metallicity core is preferentially retained. However, this effect is mild enough that differences in star formation rates are also a major (or even dominant) contributor, especially as tidal forces can either trigger or suppress the gravitational instabilities that lead to star formation.

We might assume that ram pressure would have a stronger effect than tides. Indeed it is often shown that ram pressure can produce dramatic stripping in dwarf galaxies \citep[e.g.][for a review]{2010AdAst2010E..25M}. However, comparing observations to cosmological simulations, \citet{2014MNRAS.442.1396W} found that among high-mass dwarf galaxies (stellar mass $M_s>10^{8.5}$), only $30$\% were quenched. Similarly, \citet{2018MNRAS.474.2194E} found that environmental effects did not significantly affect the width of the metal distribution function (MDF) in zoom-in simulations from the FIRE-2 and LATTE simulation suites. These simulations had a mass resolution of $\sim7000$ M$_\odot$. Ram pressure stripping of metals may be sensitive to resolution, as low-density metal-rich outflows may escape with less mixing at higher resolution. Pushing a greater fraction of metals into the wind would cause the dwarf galaxy metallicities to be more sensitive to large-scale environmental effects. These authors also only examined the total MDF of the dwarf galaxies, and did not consider the spatial distribution of the metals. 


The simulations of \citet{2013MNRAS.436.1191T} also examined tidal stripping of dwarf galaxies, concluding that AGB stars are critical for reducing the N/O ratio in interacting dwarf galaxies. However, these simulations did not include explicit chemodynamics (that is, the production and mixing of metals is not explicitly tracked in the simulations), and thus their results rely on assumptions on the metal content of galaxy winds.

The simulations of \citet{2015ApJ...815...77S} explored the effects of ram-pressure on a Magellanic Cloud model. They found that a front is formed, but that ram pressure stripping was not strong. Again, these simulations did not explicitly track metallicities, and also did not include a wind produced by star formation. Such a wind could transport metals further from the center of the dwarf galaxy potential, where they can be more easily stripped.

In this work, we perform high-resolution simulations with explicit chemodynamics from idealized initial conditions within both a tidal field and a wind tunnel based on a gas halo model. We examine the effects of tidal forces and ram pressure on the metal contents of dwarf galaxy winds and disks, and on the spatial distribution of metals within a dwarf galaxy. We focus on a dwarf galaxy with properties intermediate between the Large and Small Magellanic Clouds (section~\ref{section_galaxymodel}), within a gas halo and tidal field similar to that of the Milky Way. Simulations have shown \citep{2012MNRAS.421.2109B} that the Milky Way's interactions with the Magellanic Clouds are weak, and that the observed interaction effects are the result of the Magellanic Clouds interacting with each other. Thus, in addition to being broadly applicable to massive dwarf galaxies in general, our results will shed light on how the Magellanic Clouds might have evolved chemodynamically if they interacted only with the Milky Way and not each other.

The remainder of this paper is organized as follows. In Section~\ref{section_method} we describe our numerical method and simulation set-up. In Section~\ref{section_results}, we describe the results of these simulations. In Section~\ref{section_discussion} we compare our results to other work and discuss potential numerical issues. We then summarize our conclusions in Section~\ref{section_conclusions}.

\section{Method}\label{section_method}

\subsection{Simulation code}
We use a version of the GCD+ smoothed-particle hydrodynamics (SPH) code \citep{2003MNRAS.340..908K,2012MNRAS.420.3195B,2013MNRAS.428.1968K,2014MNRAS.438.1208K}. This code includes a stochastic star formation model that relaxes the single stellar population assumption, allowing different star particles to represent stars of different masses. Star particles return energy and metals to the ISM through supernovae and stellar winds. The Plummer-equivalent force softening length is $2$ pc. Smoothing lengths are calculated dynamically through an iterative method, so that each particle has $\approx58$ neighboring particles. The minimum smoothing length is $2$ pc, which means that particles in very dense regions have $>58$ neighbors. The metal content of particles is tracked throughout the simulation, and a sub-grid diffusion model allows metals to spread between particles. Our version of GCD+ includes modified algorithms for metal deposition and diffusion, as described in \citet[][hereafter Paper I]{2016ApJ...822...91W}, and a variable background potential to represent the varying tidal forces on a satellite galaxy moving through a host galaxy potential, as described in Paper II. In this paper, we have further extended the code to include the gas component of this host galaxy to model ram pressure, as we will describe in section~\ref{section_rammodel}.

\begin{table}
\begin{center}
\begin{tabular}{lcccccccc}
\hline\hline
Run & $M_h$ & $n_0$ & $v_0$ & $R_p$ & $f_R$ & $f_T$ & $f_{RT}$ & $P_C$\\
~ & (M$_\odot$) & cm$^{-3}$ & km s$^{-1}$& kpc & ~ & ~ & ~ & ~\\
\hline
\multicolumn{8}{l}{Circular Orbits}\\
A & $10^{12}$ & $0.460$ & $190.0$ & $100$ & $1.000$ & $1.00$ & $1.00$ & $1.00$\\
A* & $10^{12}$ & $0.460$ & $190.0$& $100$ & $1.000$ & $1.00$ & $1.00$ & $1.00$\\
B & $10^{12}$ & $0.046$ & $190.0$& $100$ & $0.100$ & $1.00$ &  $0.10$ & $0.10$\\
C & $10^{11}$ & $0.460$ & $72.8$& $100$ & $0.147$ & $0.17$ &  $0.86$ & $0.15$\\
D & $10^{11}$ & $0.046$ & $72.8$& $100$ & $0.015$ & $0.17$ & $0.09$ & $0.02$\\
\multicolumn{8}{l}{Elliptical Orbits}\\
A- & $10^{12}$ & $0.460$ & $150.0$ & $60$ &~&~&~&~\\
A-{}- & $10^{12}$ & $0.460$ & $100.0$ & $29$ &~&~&~&~\\
A-{}-{}- & $10^{12}$ & $0.460$ & $50.0$ & $11$ &~&~&~&~\\
B- & $10^{12}$ & $0.046$ & $150.0$ & $60$ &~&~&~&~\\
B-{}- & $10^{12}$ & $0.046$ & $100.0$ & $29$ &~&~&~&~\\
B-{}-{}- & $10^{12}$ & $0.046$ & $50.0$ & $11$ &~&~&~&~\\
\hline
\end{tabular}
\end{center}
\caption{\label{ictable} \textup{
Summary of the run parameters. $M_h$ is the mass of the host galaxy gravitational halo, $n_0$ is the central density of the host galaxy gas halo, $v_0$ is the initial speed of the dwarf galaxy, $R_p$ is the distance from host center at pericenter (equal to the initial radial distance of the dwarf galaxy in circular runs), $f_R$ is the magnitude of ram pressure relative to Run A. For runs with circular orbits, $f_T$ is the magnitude of tidal forces relative to run A, $f_{RT}$ is the ratio between the magnitudes of ram pressures and tidal forces relative to Run A, where a high number indicates a stronger ram pressure relative to tidal forces, and $P_C$ is the relative magnitude of thermal confinement pressure relative to Run A}.
}
\end{table}

\subsection{Galaxy model}\label{section_galaxymodel}
We use the same dwarf galaxy model as in Papers I \& II. This model consists of a disk of gas and stars within a dark matter halo, with properties similar to those of disk-like irregular or Magellanic-type galaxies such as the Magellanic Clouds. We briefly summarize the dwarf galaxy models here, but further details and motivations are provided in Papers I \& II. The total disk mass is $5\times10^8$ M$_\odot$, with a gas fraction of $f_g=0.5$. The stellar disk has a scale height of $100$ pc and a scale length of $540$ pc. The gas disk has a scale length of $860$ pc, and the vertical distribution of gas is initially set by the criterion of hydrodynamic equilibrium, although stellar feedback and radiative cooling cause the gas to rapidly move away from its initial equilibrium state. The initial metal abundances are $[\alpha/\mathrm{H}]=-2$ for all $\alpha$ species, and $[\mathrm{Fe}/\mathrm{H}]=-3$, giving $[\alpha/\mathrm{Fe}]=1$. The metallicity gradient is initially flat, and so any metallicity gradient produced in the simulations is a result of explicitly-modelled evolution.

The disk consists of $5\times10^5$ particles, giving a mass resolution of $1000$ M$_\odot$. This is placed inside an active dark matter halo with an NFW profile \citep{1997ApJ...490..493N} of mass $9.5\times10^9$ M$_\odot$ and concentration parameter $c=10$, which consists of $9.5\times10^5$ particles. 

\subsection{Tidal forces and ram pressure}\label{section_rammodel}

We perform each simulation in the center-of-mass frame of the dwarf galaxy as it orbits through the halo of a host galaxy. The tidal field of the host galaxy is calculated analytically (see Paper II). However, an analytic method is not sufficient to model ram pressure, which relies on the complex hydrodynamic interactions between the dwarf's gas and the host galaxy's gas halo. On the other hand, directly modelling the entire host gas halo as a system of particles with sufficient mass resolution to simultaneously resolve the dwarf galaxy would be prohibitively expensive, requiring billions of particles for each run. Much of this computational expense would also be unnecessary, as here we are not interested in the evolution of the host galaxy itself. To solve this issue, we place the galaxy in a cubic box of width $160$ kpc which follows the dwarf galaxy as it moves through the host halo, and only follow the evolution of material within this region. This is essentially a wind tunnel model, with gas particles entering on one end and exiting on the other.

To build this wind tunnel box, we divide the host galaxy gas halo into $100^3$ cubic cells of width $5$ kpc, where the density and temperature of the center of each cell is set by an analytic function (given below). These cells are defined in the host galaxy frame. If the motion of the $160$ kpc cubic box through this grid of cells causes the center of a cell to enter this box, the cell is populated with particles. If a cell center leaves this region, all of the gas particles it contains are deleted. This effectively produces inflow-conditions on the ``forward'' edge of the box, and outflow-conditions on the ``outward'' edge of the box. We also freeze the temperature and halo-frame velocity of particles in a boundary zone two cells thick on each wall of this box. 

We populate a cell by randomly generating particles evenly throughout the cell. The initial temperatures are linearly interpolated between the cell-centered values of the surrounding cells. We determine the temperature and density of the gas halo cells from profiles which model a Milky Way gas halo. Following \citet{2015ApJ...815...77S}, we use the $\beta$ profile parameters of \citet{2013ApJ...770..118M} for the distribution of hot gas, and use the large distance limit for the density distribution,

\begin{equation}\label{eq_ndist}
n(r) = n_0 \frac{r_c}{r}^{3\beta},
\end{equation}
where $n_0=0.46$ cm$^{-3}$, $r_c=0.35$ kpc, and $\beta=0.71$. We use this model as a base Milky Way model, but set $n_0=0.046$ cm$^{-3}$ in some runs, to isolate the evolution in a system with reduced ram pressure. We note that the scatter of halo gas masses for galaxies of similar properties was found to be quite large in the Illustris simulations \citep{2016MNRAS.462.3751K}, and hence it is not unusual for two galaxies with the same dark matter halo mass to have gas halos whose masses differ by a factor of ten.

The gravitational potential of the host galaxy is set as an NFW halo. We use two different halo masses to investigate the change in tidal forces, using $M_h=10^{12}$ M$_\odot$ as a Milky Way model, and $M_h=10^{11}$ M$_\odot$ for the lower-mass model with weak tides. We assume a fixed concentration parameter of $c=12$. The concentration parameter only varies slowly with halo mass and with a large scatter \citep{2007MNRAS.381.1450N}, and so again it is reasonable to change the halo mass by a factor of ten without changing the concentration parameter.

To set the temperature of the gas halo, we again follow \citet{2015ApJ...815...77S} and use the temperature profile of \citet{1998ApJ...497..555M},

\begin{equation}\label{eq_Tdist}
T(r) = \gamma \frac{G\mu m_p M(r)}{3rk_B}
\end{equation}
where $M(r)$ is dominated by the mass of the NFW halo, and $\gamma$ is the adiabatic index. The abundances of the halo are the solar values scaled down by $10^{-2}$.

\subsection{Simulations}

We have produced eleven runs, each with the same dwarf galaxy model, but with various host galaxy and orbital properties. Runs A, A-, A-{}-, A-{}-{}-, B, B-, B-{}-, and B-{}-{}-, have a Milky-Way mass background potential, while Runs C and D have a lower-mass background potential. Runs A (and A- etc) and C have a Milky-Way mass host galaxy gas halo, while Runs B (and B- etc) and D have a lower-mass host galaxy gas halo. We have varied these values so that we can determine the effects of tides and ram pressure in a Milky Way environment by comparing with runs where tides and ram pressure should be much weaker. In particular, both tidal forces and ram pressure should be very weak in Run D, and so this run forms a basis we can compare with the simulations with stronger tides and/or ram pressure.  The simulation properties are summarized in Table~\ref{ictable}.

We produced five runs with circular orbits to investigate where the strengths of ram pressure and tidal stripping are constant over time, to facilitate direct comparison between the runs. These runs are labelled A-D without any `-' suffix, and have a constant orbital radius of $r=100$ kpc.

We also produced six runs with elliptical orbits to investigate the more impactful effects of ram pressure and tides at low pericentres. These runs have the same host galaxy properties as the runs with circular orbits, but with a different initial velocity. These are indicated with a number of `-' signs in the suffix of the run name, where the greater the number of `-' signs, the closer the orbital pericenter is to the host galaxy centre. These runs have the same host galaxy conditions as Runs A and B. The initial velocities $v_0$ and radii at pericenter $R_p$ are given in Table~\ref{ictable}.

Additionally, we performed an additional run with identical initial conditions to Run A, except with a different random seed. This run, which we call Run A*, is used to quantify the level of variation caused by stochastic variations in the star formation rate.

The magnitude of ram pressure is proportional to $\rho v^2$, where $\rho$ is the density of the medium (proportional to $n_0$) and $v$ is the speed of the galaxy relative to the medium. For the runs with circular orbits, we use this to determine the strength of ram pressure of all runs relative to Run A, defined as $f_R = \rho_i/\rho_A (v_{0i}/v_{0A})^2$ for each run $i$.

The dwarf galaxies in the circular runs orbit at a constant distance through a potential of the same concentration parameter, and so we can characterize the magnitude of tidal forces with a single parameter  $f_T$, which can be calculated analytically from the NFW potential. If we normalize the tide strength so that $f_T=1$ for runs A and C, we find that $f_T=0.17$ in runs B and D. We can then define the ratio between the tide and ram pressure strengths relative to Run A as $f_{RT} = f_R/f_T$.

Finally, although ram pressure stripping is insensitive to gas temperature \citep[e.g.][]{2009ApJ...694..789T}, the temperature of the gas halo is important for thermal pressure {\em confinement} of the dwarf galaxy's gas. For a circular orbit, using equations \eqref{eq_ndist} and \eqref{eq_Tdist}, and assuming an ideal gas law, the pressure of the halo gas that the dwarf galaxy encounters is proportional to the product of $M_h(r=100\mathrm{~kpc})$ and $n_0$. We can calculate the relative confinement pressure $P_C$ from this, normalising the pressure so that for Run A, $P_C=1$.

All of these values are given in Table~\ref{ictable} for the circular runs to aid in the interpretation of our results.


\begin{figure*}
\begin{center}
\includegraphics[width=\textwidth]{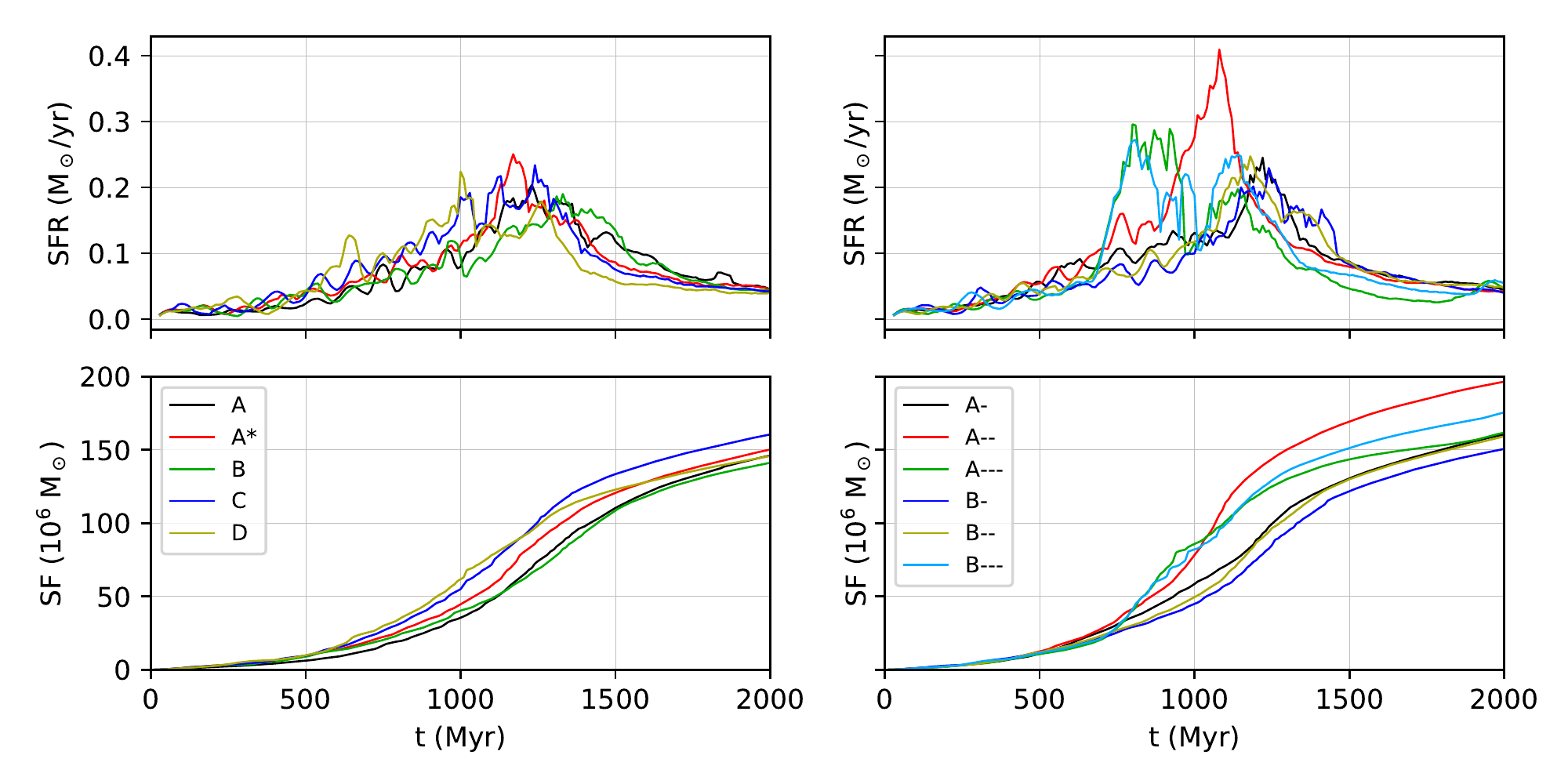}
\end{center}
\caption{\label{sfr}
Top panels: Star formation rates versus time. Bottom panels: Cumulative masses of star formation versus time.
}
\end{figure*}


\begin{figure*}
\begin{center}
\includegraphics[width=.49\textwidth]{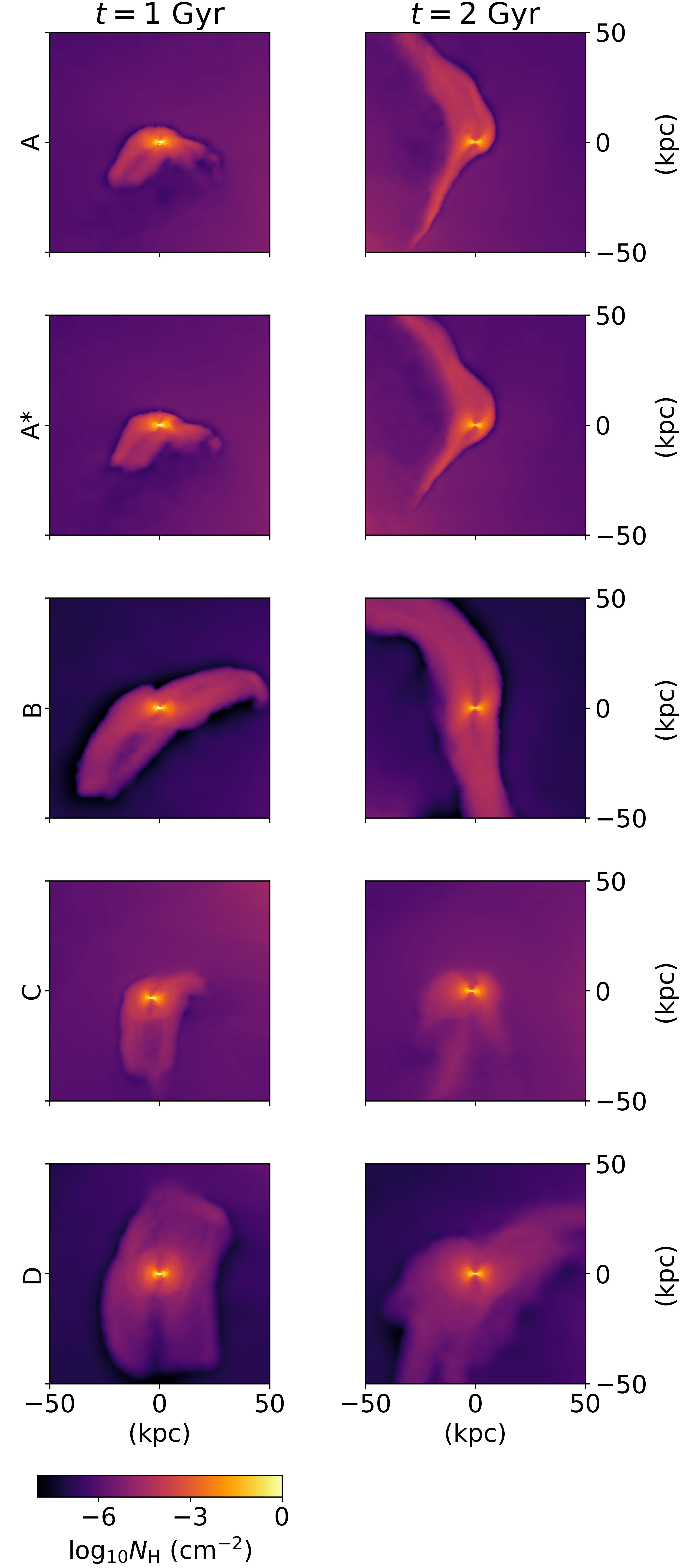}~\includegraphics[width=.49\textwidth]{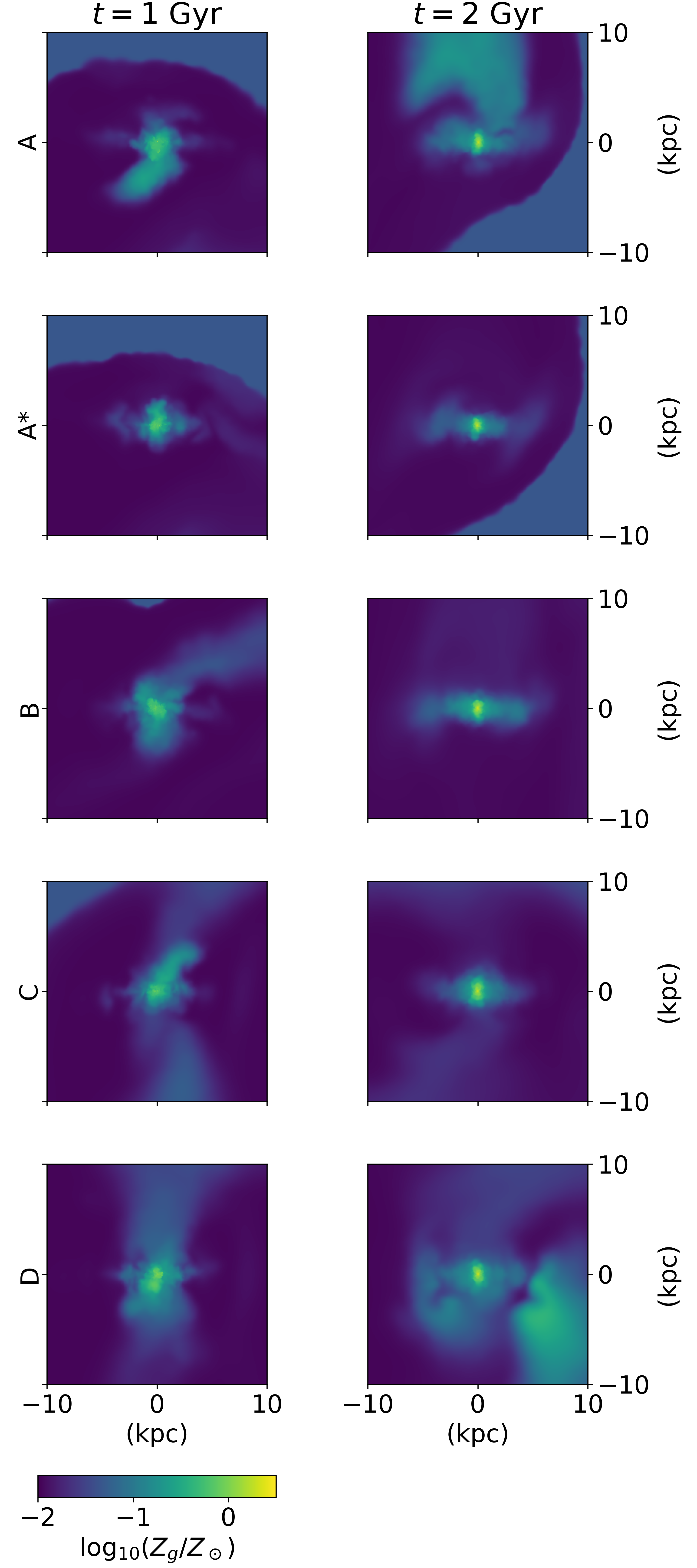}
\end{center}
\caption{\label{evolution}
Edge-on snapshots for runs with circular orbits at $t=1$ Gyr and $t=2$ Gyr. Left: column density, integrated along the line-of-sight. Right: mass-weighted mean metallicity along the line-of-sight. Note the different spatial scales of the plots.
}
\end{figure*}

\begin{figure*}
\begin{center}
\includegraphics[width=.49\textwidth]{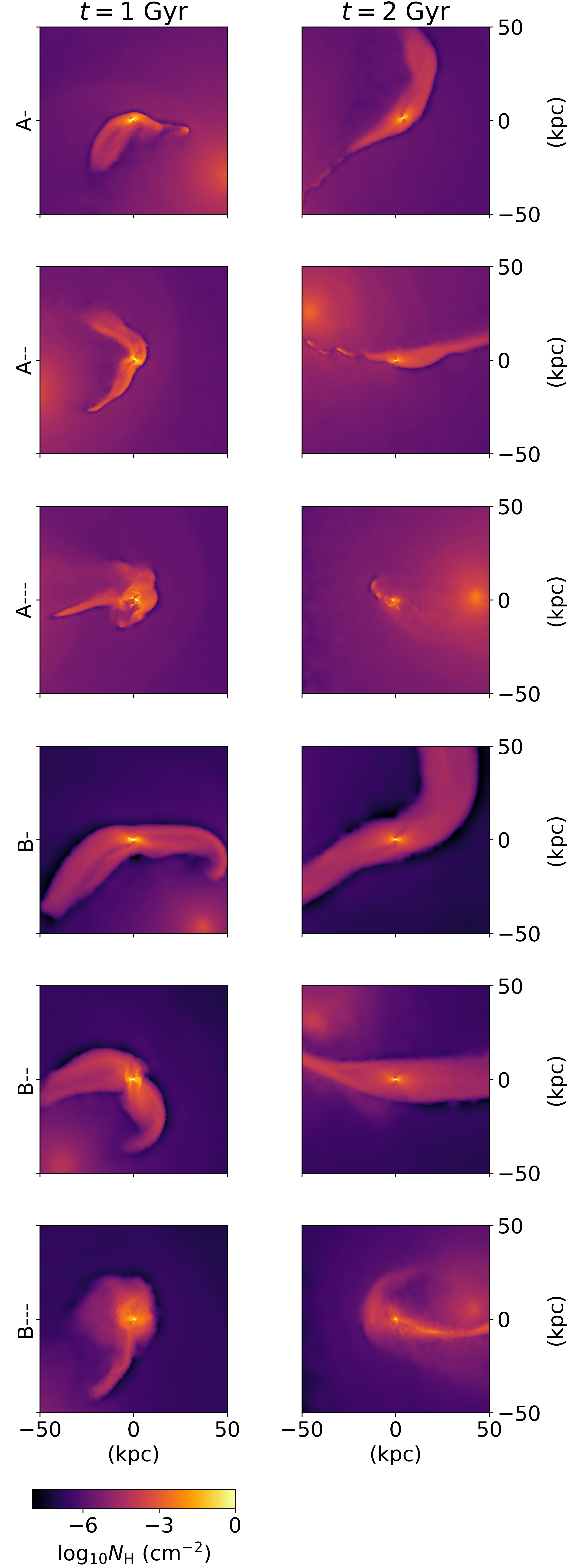}~\includegraphics[width=.49\textwidth]{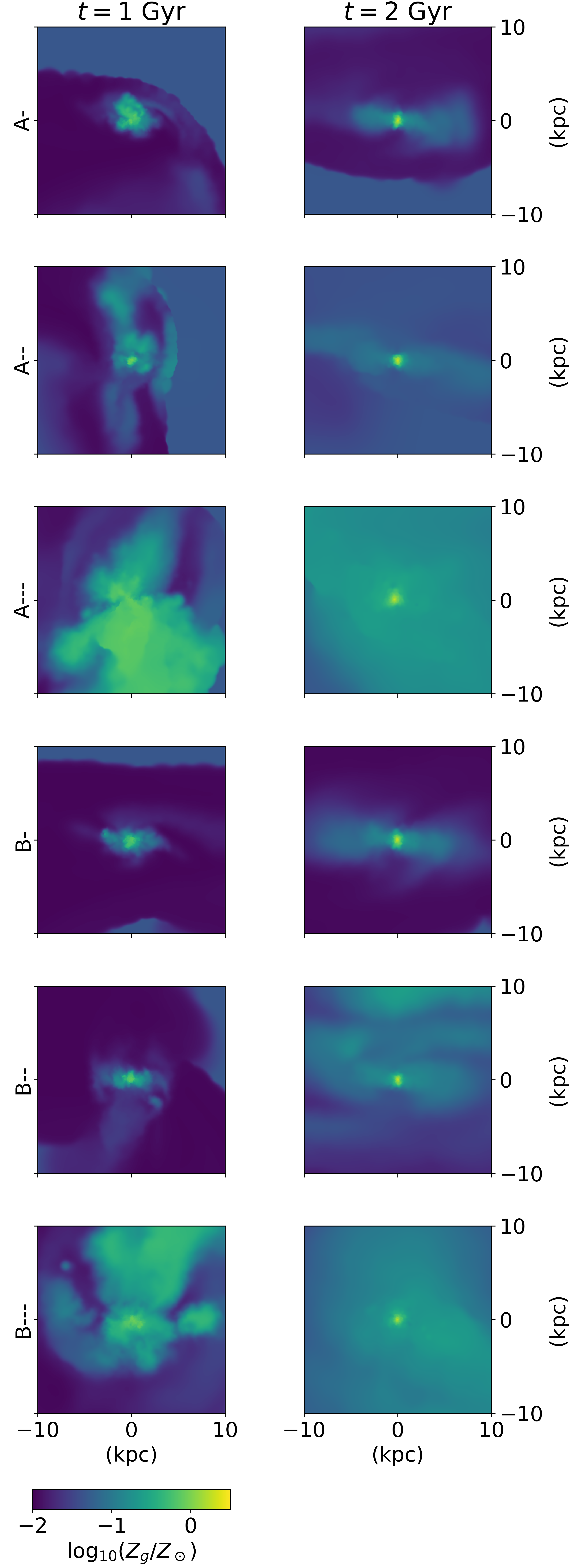}
\end{center}
\caption{\label{evolution-ellipse}
Edge-on snapshots for runs with elliptical orbits at $t=1$ Gyr and $t=2$ Gyr. Left: column density, integrated along the line-of-sight. Right: mass-weighted mean metallicity along the line-of-sight. Note the different spatial scales of the plots.
}
\end{figure*}

\begin{figure*}
\begin{center}
\includegraphics[width=\textwidth]{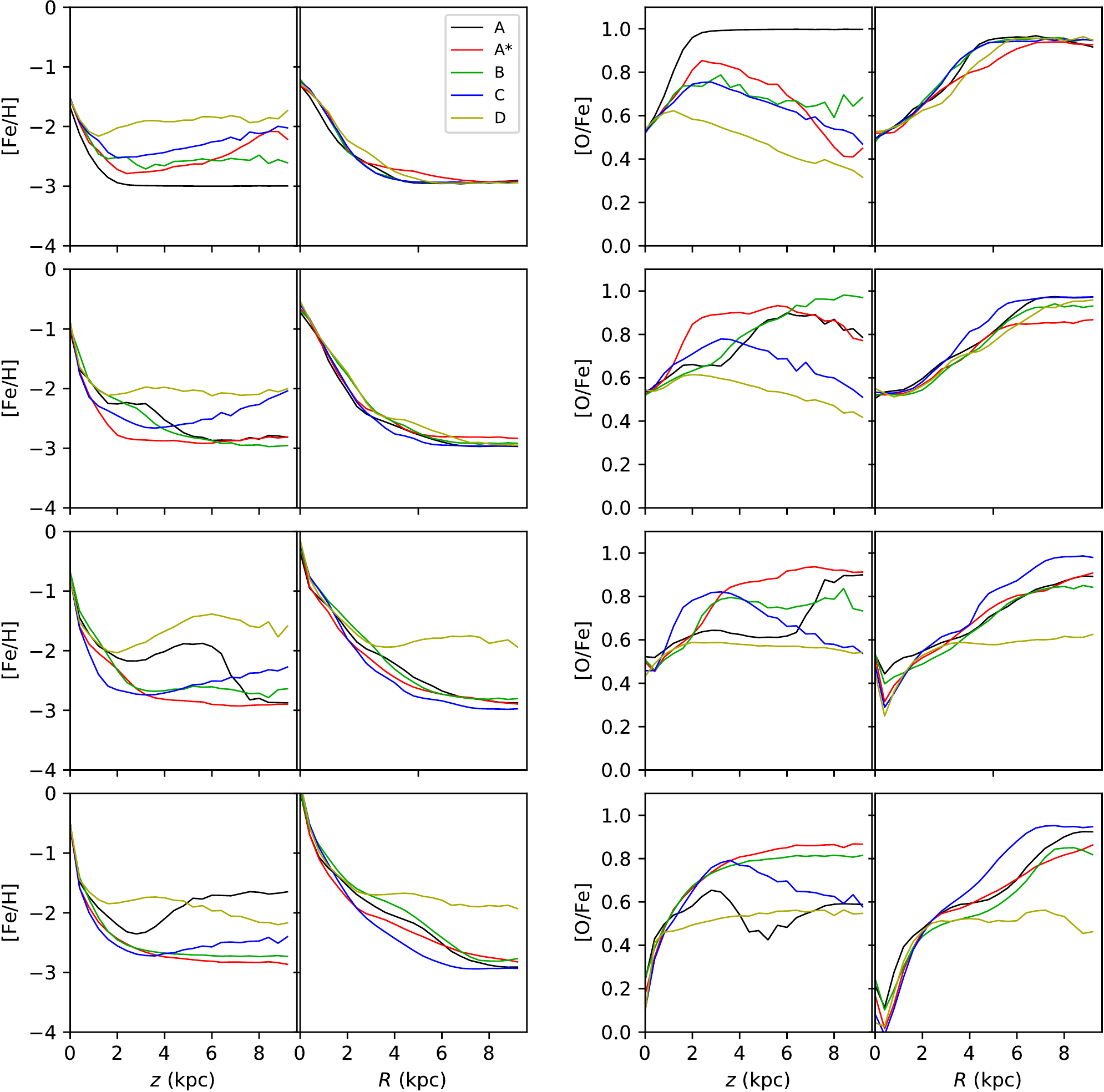}
\end{center}
\caption{\label{zslopes}
Gas-phase [Fe/H] and [O/Fe] metallicity distributions for runs in circular orbits, in vertical ($z$) and radial ($R$) directions, in intervals of $500$ Myr, from $t=500$ Myr (top) to $t=2000$ Myr (bottom).
}
\end{figure*}

\begin{figure*}
\begin{center}
\includegraphics[width=\textwidth]{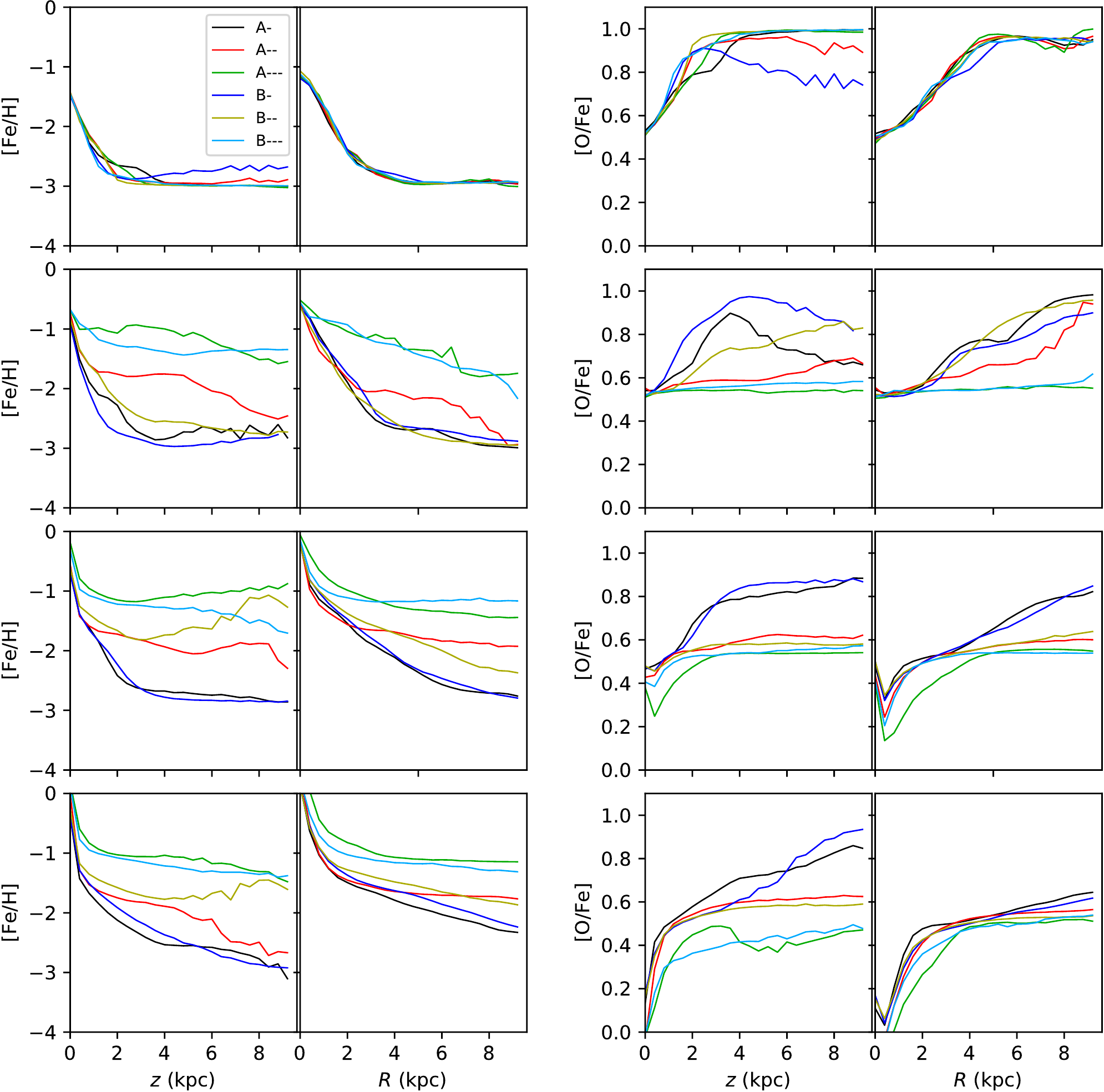}
\end{center}
\caption{\label{zslopes-ellipse}
Gas-phase [Fe/H] and [O/Fe] metallicity distributions for runs in elliptical orbits, in vertical ($z$) and radial ($R$) directions, in intervals of $500$ Myr, from $t=500$ Myr (top) to $t=2000$ Myr (bottom).
}
\end{figure*}

\begin{figure*}
\begin{center}
\includegraphics[width=\textwidth]{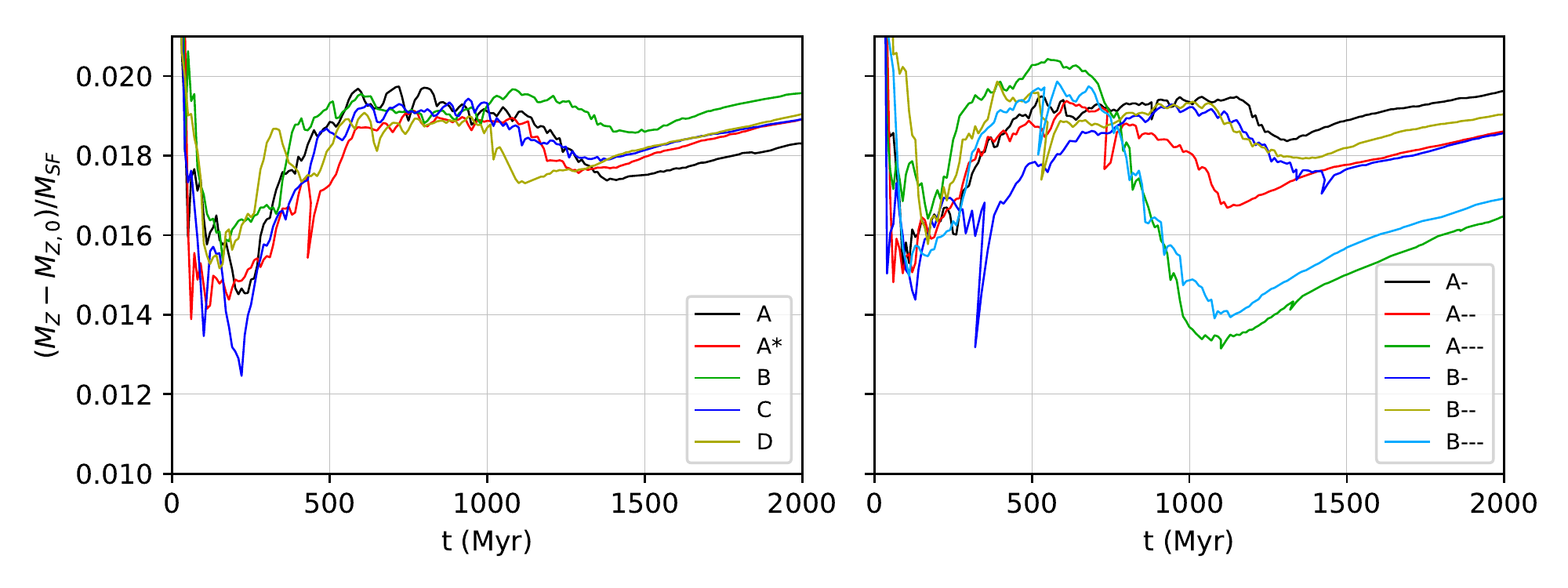}
\end{center}
\caption{\label{metevolve}
The effective yield, defined as the increase in disk ($R<5$ kpc, $|z|<2$ kpc) metal mass as a fraction of star formation mass, including both stars and gas.  Left: Runs with circular orbits. Right panel: Runs with elliptical orbits.
}
\end{figure*}

%



\begin{figure}
\begin{center}
\includegraphics[width=\columnwidth]{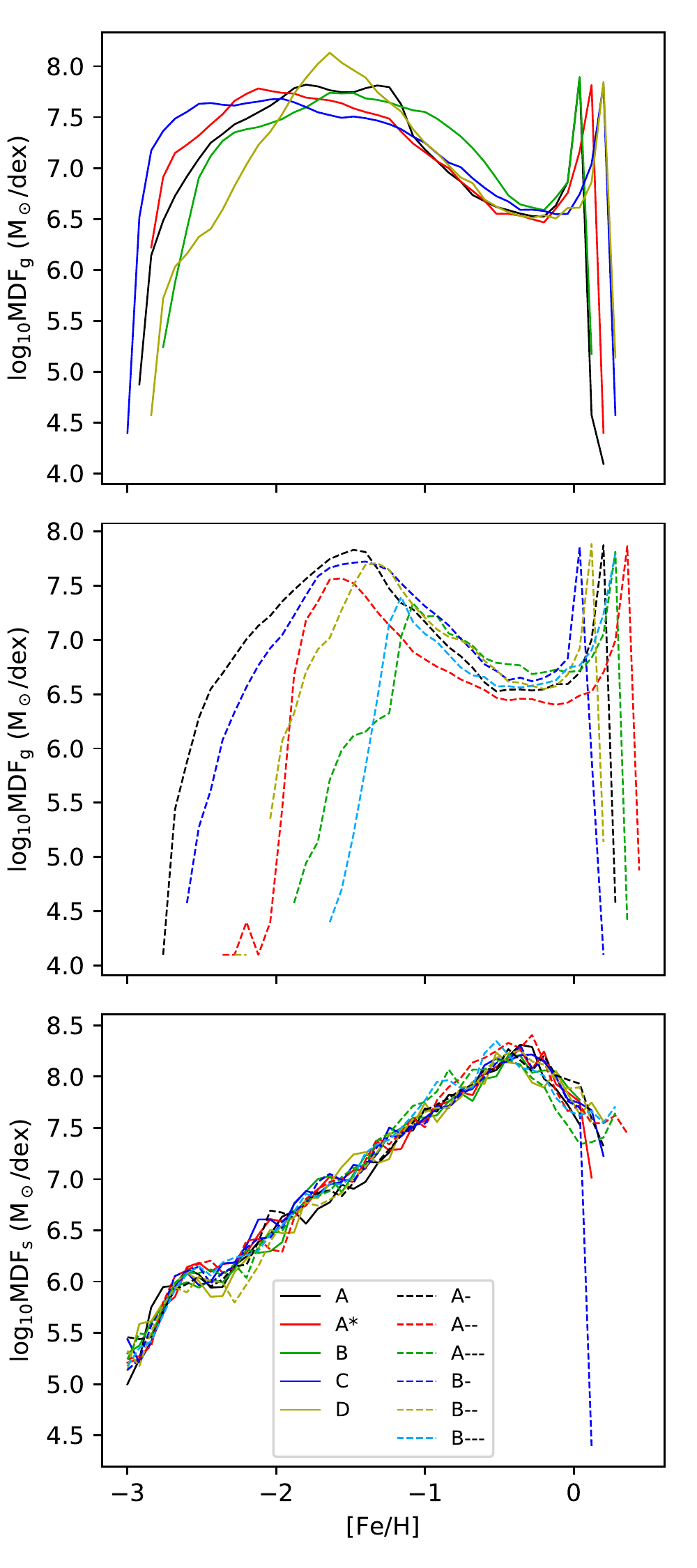}
\end{center}
\caption{\label{ramMDF}
MDFs at t = 2 Gyr, for all runs. Top panel: gas component in runs with circular orbits; middle panel: gas component in runs with elliptical orbits; bottom panel: stellar component in all runs.
}
\end{figure}

\section{Results}\label{section_results}

Figure~\ref{sfr} shows the star formation rates and cumulative star formation for all runs. The cumulative star formation is defined as the total mass of gas that has been transformed into stars by this point in time. As in Papers I \& II, at first the star formation rate gradually increases as gravitational instabilities drive gas inwards. The star formation reaches a peak rate and then starts to decline as gas is consumed.

Runs C and D have peaks that are earlier and greater in magnitude than the peaks in Runs A and B. Run C also has the greatest cumulative mass of star formation for most of the simulation time. These two runs have the weakest tidal forces, $f_T$. However, there is no monotonic trend between $f_{RT}$ or $f_R$ and the total star formation, so there is no clear connection between ram pressure and star formation here. The final star formation mass of Run D is also very close to that of Run A, despite the large difference in $f_R$. Given the chaotic nature of N-body dynamics \citep{2009MNRAS.398.1279S}, the reliance of star formation on gravitational instability, and our stochastic star formation algorithm, many of the differences in star formation between the runs with circular orbits can be attributed to psuedo-random perturbations amplified by gravitational instabilities, as in Paper II. We note that, despite having the same initial conditions, the different random seed of Run A* causes its cumulative star formation to significantly differ from Run A, with differences in cumulative star formation at any particular time often larger than the differences between Run A* and Run D. The peaks in star formation rate also occur at different times and with different magnitudes in Run A and Run A*. Hence our simulations do not show evidence that mild and constant ram pressure and tides have a direct influence on star formation, beyond being an additional source of perturbations.

The more dramatic interaction effects in the elliptical runs tell a different story. Overall, the star formation rates are generally larger, due to tidal stirring generating strong star-forming instabilities. Tides do have a significant effect here. The peaks in star formation appear near pericenter, and the runs with more circular orbits that reach pericenter later generally show peaks that occur later. Runs A-{}-{}- and B-{}-{}- have the same orbit and experience the same tidal forces, and have an initial peak of star formation at a similar time and a similar magnitude. Runs B- and A- show a similar correspondence. As the B runs have a halo gas density ten times smaller than the A runs, and should thus have ram pressure effects that are ten times weaker, these agreements shows that these bursts of star formation rate are caused by tidal stirring, and not by ram pressure compression.

The exceptions are Runs A-{}- and B-{}-, which share an orbit, but have dramatically different star formation rates -- Run A-{}- has a particularly large burst of star formation. Examining the evolution of that run in detail, we found that, solely in this run, the ram pressure, outflow rate, tidal stretching, and dwarf galaxy orbit happen to be arranged such that the outflow produced during the first $\sim800$ Myr of star formation is rapidly driven back into the galaxy as it dives into the denser parts of the halo (but before it reaches pericenter), stimulating a large burst in star formation. Again, the main effect of the halo gas is through confining the outflows, and not through ram pressure stripping, and it just happens that in this particular setup the confinement happens to drive a strong burst of star formation.

Figure~\ref{evolution} shows snapshots of the column density and metallicity of the circular orbit runs at $t=1$ Gyr and $t=2$ Gyr. Outflows are visible in the column density plots for all runs, and the effect of tides is also clear, stretching and bending the outflows. This is most visible in Runs A, A*, and B where the $M_h$ is greatest, and less visible in Runs C and D where tidal forces are smaller. The effect of the gas halo is also clear here. Rather than being a source of stripping through ram pressure, the background gas halo {\em confines} the outflows and keep the gas closer to the dwarf galaxy. Runs A and C both have more confined outflows than Runs B and D. The extent of the wind appears to depend more on the gas halo density $n_0$ than on the ram pressure strength $f_R$ or the ram-pressure-to-tides ratio $f_{RT}$. We also note that Runs B and C have similar values of $P_C$, but that the outflows are more confined in Run C where $n_0$ is higher. This suggests that the confinement is not only caused by thermal pressure, but also by the amount of mass the outflow must plough through.

However, these outflows do not appear to be efficient at transporting metals out of the dwarf galaxy. The metallicity plots in Figure~\ref{evolution} show that while these outflows have a greater metallicity than the environment of the dwarf galaxy, they are not metal-rich compared with the dwarf galaxy core, and most of the metals remain concentrated into the inner few kiloparsecs of the dwarfs. The outflows mix with and entrain low-metallicity gas, and do not consist only of metal-rich supernova ejecta. Additionally, continuous star formation in the center of the dwarf galaxy further increases the central metallicity but outflowing gas can no longer be enriched after being launched, and thus the metallicity of outflows lags behind the metallicity of the dwarf galaxy center. Hence, while interactions have a clear effect on the {\em morphology} of the outflows, this does not appear to have an impact on the metallicity of the dwarf galaxy itself, as the outflows are ineffective at removing metals from the dwarf galaxy.

More significant effects are found in the elliptical runs plotted in Figure~\ref{evolution-ellipse}. The effects of pressure confinement is clear in the density plots, as the B runs with lower gas density show more freely flowling winds than the A runs. The effects of tides are also clearly visible, particularly in Runs A-{}-{}- and B-{}-{}-, where the outflows are very disturbed. In the elliptical runs, the tidal shocks do disturb the galaxies enough to redistribute metals beyond the disk (as well as processing the dwarf disks into a more elliptical configuration in Runs A-{}-{}- and B-{}-{}-), and it appears that this effect increases as tidal forces increase. However, pressure effects only appear to be significant on the outflow at large distances, as the metal distributions in the A runs are similar to those of the B runs with corresponding orbits.

To gain more insight into the cause of these effects, we plot the metallicity profiles in Figure~\ref{zslopes}. Here we have plotted the gas-phase metallicity ratios [Fe/H] and [O/Fe] as a function of $|z|$, the distance from the disk plane (the vertical profile), and as a function of $R$, the radial distance within the disk plane (the radial profile). The vertical profile only includes gas within $R<5$ kpc, and the radial profile only includes gas within $|z|<2$ kpc. The central concentration of metals is clear by the sharp peak in [Fe/H] at low $z$ and $R$. Unlike [Fe/H],  [O/Fe], increases with distance. This is because regions of recent star formation have a greater relative abundance of iron due to Type-II supernovae. Hence we see [O/Fe] essentially follows the inverse of [Fe/H], dropping at low $z$ and $R$.

Much of the variations between runs can be explained by examining the star formation rates. The star formation is bursty, and the metal distribution depends on whether there has been a recent burst of star formation, and how dramatic it was. Dramatic bursts of star formation can produce outflows that rapidly transport metals large distances from the dwarf galaxy center, although as stated above, the metallicity of these outflows does not reach that of the star-forming galaxy center. 


Run D has the earliest bursts of star formation and hence the earliest outflows, producing a higher metallicity at $z>2$ kpc compared to other runs, as seen in Figure~\ref{zslopes}. The large burst of star formation at $t\approx1$ Gyr also pushes gas out radially along the disk plane, enriching the gas at $R>2$ kpc by $t=1.5$ Gyr and $t=2$ Gyr. Similarly, the bursts of star formation in Run A around $t=1.25$ Gyr enrich the gas at $z>2$ kpc for $t\geq1.5$ Gyr. Run B has the lowest total star formation, and has only a single burst of moderately rapid star formation, and so the metals remain more concentrated in the core.  But the dependence on stochastic star formation is most clear when we compare Run A and Run A*, where ram pressure and tidal forces should be identical, but the metallicity gradient varies greatly between the two runs, depending on the details of recent star formation.

Interestingly, Run C has a series of strong bursts of star formation and has the greatest total star formation, but the metallicity beyond $z=2$ kpc or $R=2$ kpc is among the lowest out of all the runs, especially at $t=1.5$ Gyr and $t=2$ Gyr. This can again be explained as primarily a result of star formation. In Run C, there are several bursts of rapid star formation from $t=1$ Gyr to $t=1.35$ Gyr. As is the general case in these runs, the star formation is centrally concentrated, and thus consumes the most highly enriched gas. In this case, the star formation is sufficiently rapid that this depletion is more significant than the enrichment of distant gas through outflows. We find that at $t=2$ Gyr, Run C has the lowest mass of metals in disk gas, but the greatest mass of metals in disk stars.

Confinement further contributes to this. The greater halo pressure and the greater density of the ambient medium in Run C stops winds from propagating as far as in Run D, keeping metals centrally concentrated in Run C, but allowing metals in Run D to reach higher distances. However, ram pressure {\em stripping} does not appear to have a significant effect, nor tidal stripping in the models with circular orbits.

By contrast, in the runs with elliptical orbits (Figure~\ref{zslopes-ellipse}), the effects of tides is significant. By $t=2$ Gyr, it is clear that beyond the dwarf center, [Fe/H] generally becomes higher (and [O/Fe] generally lower) with increasing tide strengths. Although there is some variation due to recent star formation, the metallicity slopes are largely paired by orbit, with no clear difference between runs with different halo gas densities. This confirms that ram pressure does not have a significant effect here, even in the inner part of the gas halo, but that tidal stripping and stirring do have a large role.

The metallicity evolution can be summarized by the evolution of the effective yield plotted in Figure~\ref{metevolve}. We define the effective yield as the increase in total disk ($R<5$ kpc, $|z|<2$ kpc) metal mass divided by the total quantity of mass consumed so far by star formation. We include the metal mass of both stars and gas in this quantity so that the effective yield only varies when metals escape the disk region. We note that this differs from other conventional definitions of the term `effective yield'. We see an early drop in the effective yield as the disk reaches equilibrium, and then all runs reach a similar yield from $0.5-1$ Gyr. After this point, the effective yield drops as bursts of star formation expel enriched gas (although as we note above, the most heavily enriched central gas is mostly retained). In the final stages ($t>1.5$ Gyr), the yield increases steadily because continuous star formation enriches the gas and produces enriched stars, but does not provide the dramatic bursts of feedback required to effectively expel gas. In this period, the rate of increase of the effective yield is similar in all simulations.

For the runs with circular orbits, the final yield at $t=2$ Gyr then largely depends on whether and to what extent the recent star formation history of the galaxy is dominated by yield-reducing dramatic bursts or by yield-increasing continuous star formation. This relates to both the timing and magnitude of the bursty phase. For example, Run B has a large final yield because its star formation is weak, producing only a small decrease in effective yield. Run A has a longer period of strong star formation, and has a much lower final effective yield. Run D also has a deeper drop in effective yield due to its bursty star formation, but this burst occurs earlier, and so continuous star formation begins earlier, with the result that it reaches a final effective yield intermediate between A and B. Run C has a similar final effective yield, despite having the most star formation, but as noted above, this results from its rapid star formation consuming high-metallicity gas and thus reducing the gas supply for outflows. This is most plainly seen by the fact that the final yield of Run A* is closer to that of Run D than Run A, despite Runs A and A* having identical orbital and halo properties. Again, for the runs with circular orbits there is no clear correlation between the final yield and the strength of tidal stripping or ram pressure stripping -- the yields appear to be dominated by stochastic variations in the star formation rate.

The runs with elliptical orbits do show environmental effects on the effective yield, though not always dominating over star formation effects. The two runs with the greatest tidal forces -- A-{}-{}- and B-{}-{}- -- show significantly lower effective yields than all other runs. However, the rest of the elliptical runs show no more variation in final effective yield than the circular runs. This is because the total mass of metals is dominated by the large mass of metals in the central region, which is often not strongly effected by environmental effects.

We can disentangle the role of the high-metallicity region by examining the metallicity distribution functions (MDF). The MDFs for the disk gas and formed stars at $t=2$ Gyr is plotted in Figure~\ref{ramMDF}. The stellar MDFs agree between all runs, showing no significant sensitivity on environmental effects. This is because stars are generally formed in the center of the galaxy, where environmental effects are at their weakest.

The gas MDFs do show an environmental dependence. The metallicities of the highest metallicity peaks vary between the runs, but as stated above these peaks represent gas within a very small central region ($\lesssim200$ pc), and are not strongly affected by environmental effects. Instead, it is the low-metallicity gas that shows evidence of tidal stripping. The elliptical runs are again paired by orbit rather than gas halo density, with a lower quantity of low-metallicity gas as tidal strength increases, independent of the density of the gas halo. This shows that tidal stripping is again dominant over ram pressure. 

There may also be an indication of a pressure confinement effect in the circular orbit runs. The runs with greater halo densities (A,A*,C) retain higher quantities of low-metallicity gas than the runs with lower halo densities (B,D). However, this effect is not large compared to the difference between Run A* and Run A, and may not be significant.


\section{Discussion \& Comparison with other work}\label{section_discussion}

The effects of interactions on dwarf galaxies have been investigated in cosmological simulations, with work that is typically focused on the role of interactions in producing a population of quenched red dwarf galaxies. These simulations typically have mass resolutions of $10^4-10^6$ M$_\odot$, and can only resolve the most massive dwarfs, typically with $M_s>10^8$ M$_\odot$. Both the Illustris simulation and semi-analytic models based on the Millennium simulations produce quenching that is too rapid compared with SDSS observations \citep{2014MNRAS.442.1363W,2015MNRAS.447L...6S} despite these simulations being performed with very different numerical methods -- the Millennium simulations were performed with the SPH code Gadget \citep{2005MNRAS.364.1105S} and the Illustris simulation was performed with the moving-mesh code AREPO \citep{2010MNRAS.401..791S}. However, without a semi-analytic model to post-process the simulation, it has been reported that the Millennium simulation produces `surprisingly inefficient' quenching \citep{2014MNRAS.442.1396W}. This particular study only considered dwarf galaxies with a gas greater than $10^{8.5}$ M$_\odot$, that are perhaps too massive to be quenched by ram pressure and tidal stripping. A study of metal stripping and star formation of more massive satellites ($M>10^9$ M$_\odot$ in Illustris \citep{2016ApJ...822..107G} found that star formation rates are concentrated and disks are truncated (but not rapidly quenched) by interactions, producing a higher observed metallicity. However, inefficient interaction effects (ram-pressure in particular) has also been found in a high-resolution simulation of the lower-mass dwarf galaxy Leo T \citep{2016ApJ...826..148E}. As for an intermediate-mass galaxy such as the LMC, truncation rather than dramatic tidal stripping is also found in simulation \citep{2005MNRAS.363..509M}. The zoom-in simulations of \citet{2018MNRAS.478..548S} investigated the effect of dwarf mass at higher resolution, and found a steep trend in quenching from a $90$\% quenching fraction for $M_s=10^6$ M$_\odot$ dwarfs down to $30$\% for $M_s=10^8$ M$_\odot$ dwarfs. In the EAGLE simulation, it was found that only $10$\% of LMC mass galaxies in Milky Way mass haloes were quenched \citep{2018MNRAS.479..284S}. The general consensus appears to be that the effects of quenching and ram pressure on Magellanic Cloud mass galaxies is small, and this is consistent with our results.

However, ram pressure is a phenomenon that can be very sensitive to numerical methodologies. This was most famously demonstrated in the `blob test' of \citet{2007MNRAS.380..963A}, where a sphere of dense cold fluid is impacted by a high-velocity hot low-density medium -- analogous to molecular gas in a satellite galaxy being stripped by a hot host halo. It was found that traditional SPH methods preserved the stability of the blob far longer than grid-based methods. This was explained by the improper calculation of pressure gradients at boundaries causing a surface-tension effect, along with a small contribution from artificial viscosity. This led to the development of new SPH paradigms, such as the pressure-entropy paradigm \citep[e.g.][]{2013MNRAS.428.2840H}. This form shows a greater agreement with grid codes, and is the form used in our version of GCD+ \citep[for numerical tests of this code, see][]{2013MNRAS.428.1968K}. Moving-mesh or `mesh-free' methods \citep{2010MNRAS.401..791S,2015MNRAS.450...53H} have also gained attention in recent years. However, comparisons between hydrodynamic methods in galaxy models and cosmological simulations have found little dependence on hydrodynamic method \citep{2012MNRAS.423.1726S,2015MNRAS.450...53H,2015MNRAS.454.2277S}, other than traditional SPH methods producing spurious over-densities and higher star-formation rates. In general, it is found that the variations between simulations is utterly dominated by the variations in sub-grid models, and not in hydrodynamics methods. Hence, as we are using a `modern' SPH method, we can be reasonably confident that our results are not affected by numerical problems with our hydrodynamics implementation.
 
\section{Summary \& Conclusions}\label{section_conclusions}

To investigate the effects of ram pressure and tidal stripping on dwarf galaxies, we have performed chemodynamical simulations of dwarf galaxies with similar properties to the Magellanic Clouds orbiting within host galaxies with dark matter and gas halos similar to the Milky Way, investigating circular and elliptical orbits, and varying both the gas and dark matter content separately to have either the Milky Way mass or one tenth the mass of the Milky Way. We have found that the effects of ram pressure on the metallicity of the dwarf galaxy disks are not significant, and that the differences between the models can be explained by stochastic variations in the star formation rate, and by the effects of tides, which are only significant at low pericenters. Ram pressure and tidal forces do affect the morphology and metallicity of outflows at large distances from the dwarf galaxy, even on dwarf galaxies circular orbits at large distances, but this does not significantly affect the metallicities of the dwarf galaxy disk if the tides are not very strong. We do find that tidal effects can truncate the dwarf galaxy in plunging orbits, but instead of lowering the metallicity of the dwarf galaxies, the effect is to remove low-metallicity outer gas. We also find that the host galaxy's halo pressure can {\em confine} outflows rather than strip them, slightly enhancing the dwarf galaxy's metallicity.

This raises two important issues. Firstly, given that our model dwarf galaxy is essentially a lone Magellanic Cloud, the lack of strong interaction effects in our simulations supports the conclusions of \citet{2012MNRAS.421.2109B} that the interaction effects observed in the Magellanic Clouds are primarily the result of interactions between the two Clouds, and not of those between the Clouds and the Milky Way -- although we have concentrated on metallicity distributions rather than on morphology.

Secondly, and more critically, how are the observed low metallicities and low gas fractions of dwarf galaxies generated if ram pressure stripping is ineffective? A clue here may be the surprisingly low metallicities of our outflows. While we use a modern hydrodynamics scheme, the wind could still be sensitive to resolution. Although our mass resolution of $1000$ M$_\odot$ is finer than that of e.g. \citet{2018MNRAS.474.2194E}, it may be that high-metallicity outflows are still suppressed at current resolutions. We plan to produce higher-resolution simulations in a future paper.

\section*{Acknowledgements}
This research was supported by the Canada Research Chair program and NSERC. The authors acknowledge the use of the {\em Guillimin} and {\em Colosse} computing clusters supported by Calcul-Qu\'{e}bec/Compute-Canada, as well as the IRIDIS High Performance Computing Facility supported by the University of Southampton, in the completion of this work. DW is supported by European Research Council Starting Grant ERC-StG-677117 DUST-IN-THE-WIND. We thank our anonymous reviewer for advice that improved the content and presentation of this paper.\\

\bibliographystyle{apj}
\bibliography{ram}

\end{document}